\documentclass[12pt]{article}
\usepackage{physics}
\usepackage{epsf}
\usepackage{aas_macros}
\usepackage{amssymb}
\usepackage{comment}
\usepackage{amsmath}
\usepackage{braket}
\usepackage{mathrsfs}
\usepackage{mathtools}
\usepackage{array}
\usepackage{fancyhdr}
\usepackage[dvipdfmx]{graphicx}
\usepackage{color}
\usepackage{cite}
\usepackage{bm}
\usepackage{url}
\usepackage{fancyhdr}
\usepackage[colorlinks,citecolor=blue]{hyperref}
\usepackage[hang,small,bf]{caption}
\usepackage[subrefformat=parens]{subcaption}
\renewcommand{\thefootnote}{\#\arabic{footnote}}

\renewcommand{\thefootnote}{\fnsymbol{footnote}}
\def\thefootnote{\fnsymbol{footnote}}

\makeatletter

\@addtoreset{equation}{section}
\makeatother

\def\be{\begin{equation}}
\def\ee{\end{equation}}
\def\ben{\begin{eqnarray}}
\def\een{\end{eqnarray}}

\usepackage{scalerel}
\usepackage[usestackEOL]{stackengine}

\def\dashint{\,\ThisStyle{\ensurestackMath{%
            \stackinset{c}{.2\LMpt}{c}{.5\LMpt}{\SavedStyle-}{\SavedStyle\phantom{\int}}}%
        \setbox0=\hbox{$\SavedStyle\int\,$}\kern-\wd0}\int}

\begin{document}


\begin{center}

\vskip .75in

{\Large \bf Kramers-Kronig relation in gravitational lensing}

\vskip .75in

{\large
So Tanaka$\,^1$,  Teruaki Suyama$\,^1$
}

\vskip 0.25in

{\em
$^{1}$Department of Physics, Tokyo Institute of Technology, 2-12-1 Ookayama, Meguro-ku,
Tokyo 152-8551, Japan
}

\end{center}
\vskip .5in

\begin{abstract}
The Kramers-Kronig relation is a well-known relation, 
especially in the field of optics. 
The key to this relation is the causality that output comes only after input. 
We first show that gravitational lensing obeys the causality in the sense 
that (electromagnetic/gravitational) waves emitted from the source arrive 
at an observer only after the arrival of the signal in geometrical optics. 
This is done by extending the previous work which is based on the thin lens approximation. 
We then derive the Kramers-Kronig relation in gravitational lensing, 
as the relation between real and imaginary parts of the amplification factor, 
which is the amplitude ratio of the lensed wave to the unlensed wave. 
As a byproduct, we find a new relation that equates integration of the 
square of the real part of the amplification factor over frequency to that 
for the imaginary part of the amplification factor. 
We also obtain a sum rule which relates the integral of the imaginary part of the amplification factor with the magnification of the first arrival image 
in geometrical optics. 
Finally, we argue that an incorrect separation of the observed 
gravitational waveform into the amplification factor and the unlensed 
waveform generically leads to the violation of the Kramers-Kronig relation. 
Our work suggests that examining the violation of the Kramers-Kronig relation 
may be used for correctly extracting the lensing signal
in the gravitational wave observations.
\end{abstract}

\renewcommand{\thepage}{\arabic{page}}
\setcounter{page}{1}
\renewcommand{\thefootnote}{\#\arabic{footnote}}
\setcounter{footnote}{0}

\section{Introduction}
Light passing through a gravitational field is bent. This phenomenon is known as gravitational lensing (GL)\cite{Bartelmann:2010fz, schneider1999gravitational}, and gravitational waves (GWs) are also subject to this effect\cite{misner2017gravitation, Bailes:2021tot}. 
One of the prominent features of GWs over light is their long wavelength nature.
Because of this, in some cases, geometrical optics which only deals with null 
geodesics breaks down and the propagation of GWs should be described by wave optics \cite{Nakamura:1999uwi,Takahashi:2003ix}.
In the regime of wave optics, the lensed GWs carry more information about a lens object since they pass through a more extended region due to diffraction\cite{ruffa1999gravitational}.

Recently, there is a discussion about the arrival time difference between 
light and GWs due to GL\cite{Takahashi:2016jom, Suyama:2020lbf, Ezquiaga:2020spg}. 
It was argued in \cite{Suyama:2020lbf} that GWs never arrive earlier than 
light if they depart from the source at the same time. 
This can be rephrased as that GL signal in wave optics regime comes only 
after the signal in geometrical optics.
This fact gives us inspiration that the Kramers-Kronig (K-K) relation,
which is satisfied as long as any system under consideration
satisfies the causality condition that output comes only after input,
also holds in GL.

The K-K relation is directly derived from causality, and actually, 
it is the relation between real and imaginary parts of a response function\cite{nussenzveig1972causality}. It is often used in the field of optics, for example, as a relation between the refractive index and the extinction coefficient. The typical application is optical data inversion\cite{lucarini2005kramers}: we can obtain data of the refractive index from that of the extinction coefficient, or vice versa. 
However, to the best of our knowledge,
the K-K relation has never been discussed in the context of GL.
This observation is sufficient to motivate us to clarify the K-K relation 
in GL and to investigate potential applications to the observations of GL of GWs.

In this paper, we first revisit the causality of GL. 
While the previous study used the so-called thin-lens approximation\cite{Suyama:2005mx} to show the causality, 
we provide an explicit proof that the causality holds true without resorting to
the thin-lens approximation. 
In Sec.~\ref{sec:KK-relation}, we derive the K-K relation in GL which gives
a non-trivial relation between the real part of 
and the imaginary part of the amplification factor. 
We also derive some relations which directly follow from the K-K relation. 
In Sec.~\ref{sec:application}, we argue that an incorrect separation of the 
observed gravitational waveform into the amplification factor and the unlensed 
waveform generically leads to the violation of the Kramers-Kronig relation. 
Given that it is observationally challenging to discern a lensing effect from a characteristic of a source\cite{Kim:2023scq},
examining the violation of the Kramers-Kronig relation has a potential to 
correctly extract the lensing signal
in the gravitational wave observations.

\section{Causality of gravitational lensing}
\label{causality}
\begin{figure}
    \centering
    \includegraphics[scale=0.5]{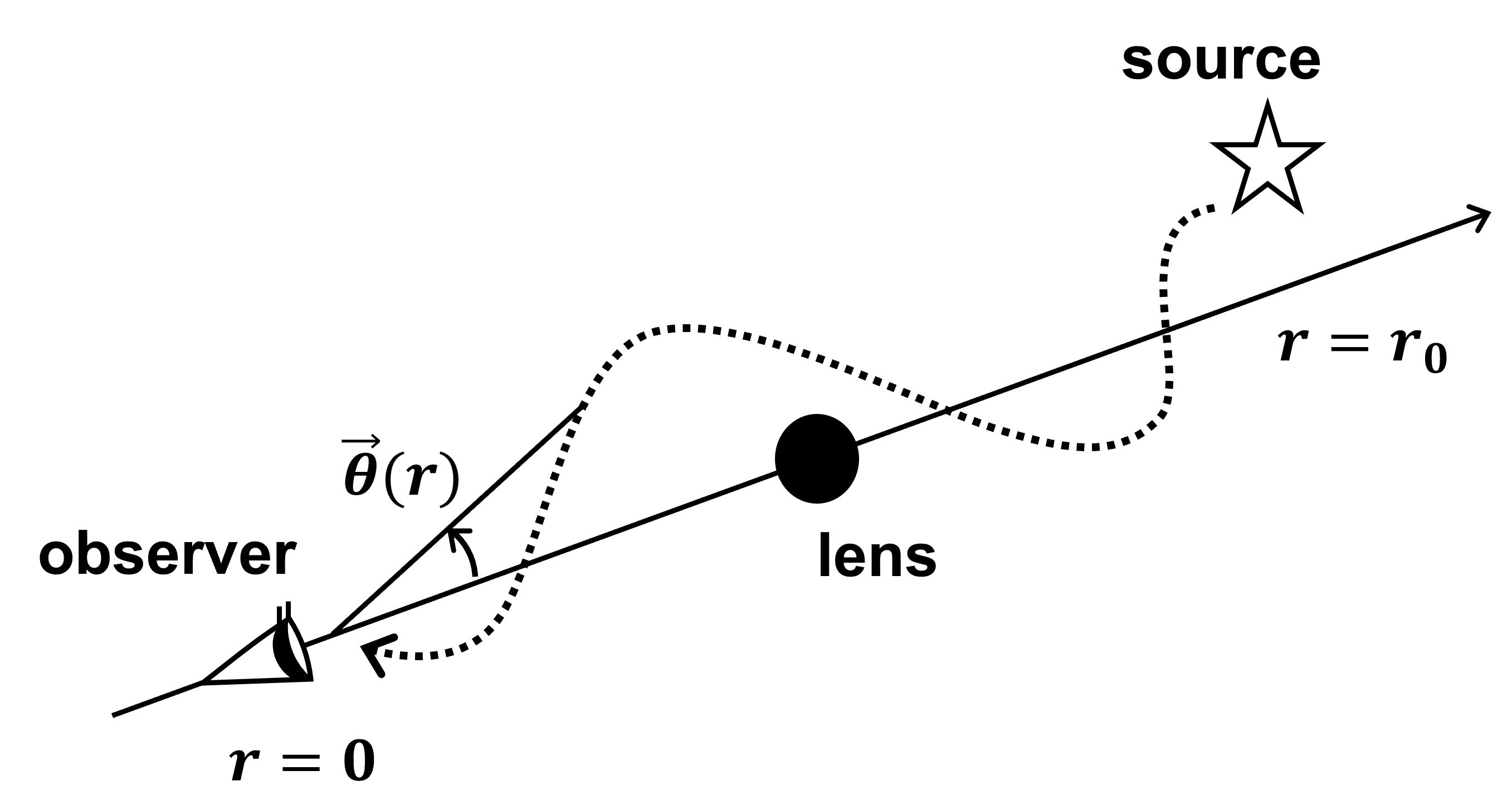}
    \caption{Schematic picture of GL. The dotted line represents a path of the waves, and all paths contribute to the path integral.}
    \label{gl}
\end{figure}

In this section, we investigate the causality of GL, which is needed to derive the K-K relation. 
Propagating waves are either GWs or electromagnetic waves both of which have
polarization degrees and the waves can be written as a product of wave 
amplitude $\phi$ and polarization vector/tensor.
The change of the polarization vector/tensor due to GL is suppressed by the gravitational potential $(\ll 1)$ \cite{schneider1999gravitational} 
and we ignore the polarization in this paper.  
The background metric $g_{\mu \nu}^B$ on which the wave propagates is given by
\be
ds^2=g_{\mu \nu}^Bdx^\mu dx^\nu=-(1+2\Phi) dt^2+ (1-2\Phi )d{\bm x}^2,
\ee
where $\Phi$ is the gravitational potential of the lensing objects
\footnote{
Here we ignore the expansion of the Universe because it is 
not important in this discussion
(see Ref. \cite{Takahashi:2005ug}).
}.
The wave equation for $\phi$ is that for a massless scalar field \cite{peters1974index}
\be
\label{causality4}
\partial_\mu ( \sqrt{-g^B} g_B^{\mu \nu} \partial_\nu \phi)=0.
\ee
The lensed wave $\phi_{L}$ is a solution of this equation.
To represent the effects of GL, it is customary to move to the frequency domain 
where the lensed waveform is simply given a product of the unlensed waveform
and the amplification factor $F(\omega)$:
\be
\phi_{L}(\omega)=F(\omega)\phi_{0}(\omega),\label{causality2}
\ee
where $\phi_{L}(\omega)$/$\phi_{0}(\omega)$ are the lensed/unlensed wave
in the frequency domain, evaluated at the observer's position. 
Then Eq.~(\ref{causality4}) can be solved in terms of $F$ and the formal solution is given by the path integral\cite{Nakamura:1999uwi}:
\footnote{
In fact, this solution is derived under the eikonal approximation and $\theta\ll 1$ \cite{Nakamura:1999uwi}.
}
\be
F(\omega)=\int\mathcal{D}\bm{\theta}(r)e^{i\omega T[\bm{\theta}]},\label{Causality5}
\ee
where
\be
T[\bm{\theta}]=\int_{0}^{r_0} dr\qty[\frac{1}{2}r^2\qty(\frac{d\bm{\theta}}{dr})^2-\Phi(r,\bm{\theta}(r))],\label{Causality6}
\ee
and $\bm{\theta}$ is a two-dimensional angular vector perpendicular to the line of sight, $r$ is a radial coordinate along the line of sight (the observer is located at $r=0$ and the source is at $r=r_0$) (see Fig. \ref{gl}). 
The time dependence of $\Phi$, which can arise when the source varies in time, 
is encoded in $r$ dependence through 
$r-t={\rm const.}$.
The first term of Eq. (\ref{Causality6}) represents the deviation of a path from the straight line, while the second term represents the time delay caused by the gravitational potential.

What we want to compute is the Fourier transform of $F(\omega)$, and to do so we first discretize the path integral. We divide the distance to the source into N parts and define
\ben
r_j&\equiv&j\Delta r,\\
\bm{\theta}_j&\equiv&\bm{\theta}(r_j),\\
\Phi_j&\equiv&\Phi(r_j,\bm{\theta}(r_j)),\\
T_j&\equiv&\Delta r\qty[\frac{1}{2}r_j^2\qty(\frac{\bm{\theta}_{j+1}-\bm{\theta}_j}{\Delta r})^2-\Phi_j],
\een
where $\Delta r=r_0/N$. Then we get
\ben
F(\omega)&=&\int\qty(\prod_{j=1}^{N-1} N_j d^2\theta_j)\exp(i\omega \sum_{j=1}^{N-1}T_j)\nonumber \\
&=&\int \prod_{j=1}^{N-1}d^2\theta_j F_j,
\een
where
\be
F_{j}\equiv N_{j}e^{i\omega T_{j}},
\ee
and 
\be
N_j=\frac{\omega r_j^2}{2\pi i \Delta r}
\ee
is the normalization factor required for F=1 in the absence of the gravitational potential. The Fourier transform of $F_j(\omega)$ is
\ben
f_j(t)&=&\int_{-\infty}^{\infty}\frac{d\omega}{2\pi}F_j(\omega)e^{-i\omega t}\nonumber\\
&=&\frac{r_j^2}{2\pi\Delta r}\frac{d}{dt}\delta(t-T_j),
\een
hence that of $F(\omega)$ becomes
\ben
f(t)&=&\int_{-\infty}^{\infty}\frac{d\omega}{2\pi}F(\omega)e^{-i\omega t}\nonumber\\
&=&\int\qty(\prod_j d^2\theta_j)\int_{-\infty}^{\infty} dt_2\cdots dt_{N-1}f_1(t-t_2)\cdots f_{N-2}(t_{N-2} - t_{N-1})f_{N-1}(t_{N-1})\nonumber\\
&=&\int\qty(\prod_j\frac{r_j^2}{2\pi i \Delta r}d^2\theta_j)\int_{-\infty}^{\infty} dt_2\cdots dt_{N-1}\nonumber\\
&&\quad\frac{d}{dt}\delta(t-t_2-T_1)\cdots\frac{d}{dt_{N-2}}\delta(t_{N-2}-t_{N-1}-T_{N-2})\frac{d}{dt_{N-1}}\delta(t_{N-1}-T_{N-1})\nonumber\\
&=&\int\qty(\prod_j\frac{r_j^2}{2\pi i \Delta r}d^2\theta_j)\frac{d^{N-1}}{dt^{N-1}}\delta(t-\sum_j T_j),\label{causality1}
\een
where we have used the fact that the Fourier transform of a product 
becomes a convolution integral and the properties of the delta function:
\be
g(t')\frac{d^n}{dt^{n}}\delta(t-t')
= g(t')(-1)^n\frac{d^n}{dt'^{n}}\delta(t'-t)
=\delta(t'-t)\frac{d^n}{dt^{n}}g(t).
\ee
Taking $N\rightarrow\infty$ limit, Eq. (\ref{causality1}) yields
\be
f(t)\propto\frac{d^{\infty}}{dt^{\infty}}\int\mathcal{D}\bm{\theta}~\delta\qty(t-T[\bm{\theta}]),
\ee
thus we conclude that
\be
f(t)=0\quad \rm if\it\quad t<T_{min}\equiv\underset{\bm{\theta}}{\min} \{T[\bm{\theta}]\}\label{causality3}.
\ee
This means that there is no lensing signal before $t=T_{min}$. Since $T_{min}$ is the time delay in the geometric optics limit, 
we can conclude that GWs never arrive earlier than light if these are emitted at the same time. 
This is the causality of GL which is crucial to prove the Kramers-Kronig
relation for the amplification factor in the next section.
As we mentioned before, Ref. \cite{Suyama:2020lbf} has shown the same result with the thin lens approximation. Thus our proof is the generalization of that.

\section{Kramers-Kronig relation}
\label{sec:KK-relation}
\subsection{Derivation}
In this subsection, we derive the K-K relation in GL. All that is needed for this is the causality and the asymptotic behavior of $F(\omega)$\cite{nussenzveig1972causality}. 
First, let us verify the analytic behavior of $F(\omega)$ that is related to the causality. From Eq. (\ref{causality3}), $F(\omega)$ can be written as
\be
F(\omega)=\int_{T_{min}}^{\infty}dt~f(t)e^{i\omega t},
\ee
or
\be
F_{ph}(\omega)\equiv F(\omega)e^{-i\omega T_{min}}=\int_{0}^{\infty}dt~ f(t+T_{min})e^{i\omega t}.
\ee
Then $F_{ph}(\omega)$ can be analytically continued to the upper half of the complex $\omega$-plane (we shall write $I_{+}$):
\be
F_{ph}(u + iv) = \int_{0}^{\infty}dt~ f(t+T_{min})e^{iut}e^{-vt},
\ee
where $v>0$.
If we assume that $F_{ph}(\omega)$ does not have any poles on the real axis (this is physically reasonable), then $F_{ph}(\omega)$ is also regular in $I_{+}$, because the term $e^{-vt}$ only improves the convergence of the integral. Furthermore, $F_{ph}(\omega)$ has its physical meaning. 
Since the time delay $T_{min}$ itself is not directly measurable, 
it is sensible to remove this degree of freedom and to use $F_{ph}(\omega)$ 
rather than $F(\omega)$. 
From now on, we focus on $F_{ph}(\omega)$ and use $F(\omega)$
for $F_{ph}(\omega)$.

Besides, we have to know the asymptotic behavior of $F(\omega)$. In $\omega\rightarrow\infty$ limit, excepting some special cases\footnote{In the case of the point mass lens with the impact parameter $y=0$, $F(\omega)$ diverges as $F(\omega)\propto\sqrt{\omega}$. However, in this case Eq. (\ref{K-K1}) also holds because $|G(\omega)|\rightarrow 0$.}, $F(\omega)$ does not diverge:
\be
|F(\omega)| < C,
\ee
where $C$ is some constant. On the other hand, in $\omega\rightarrow0$ limit, $F(0)=1$ because waves with extremely long wavelengths do not feel the gravitational field.
This can also be understood from Eq.~(\ref{causality4}). 
When $\omega=0$, any time derivatives in Eq.~(\ref{causality4}) disappear
and the wave equation coincides with the free propagation equation, 
thus $\phi_{L}(0)=\phi_0(0)$.

From above, if we define $G(\omega)$ by
\be
G(\omega)\equiv \frac{F(\omega)-1}{\omega},
\ee
then $G(\omega)$ has no poles on the real axis
\footnote{It is possible for $G(\omega)$ to diverge at $\omega=0$, for example, in the case that $F(\omega)$ contains a term like $\omega\ln\omega$. However, even in this case, there is no divergence on $\Gamma$ in Fig. \ref{Gamma} since $|G(\omega)|<1/\omega$ ($\omega\rightarrow0$) and improper integral converges.} 
and in $I_+$, and in $\omega\rightarrow\infty$ limit $|G(\omega)|\rightarrow 0$.
Therefore, by using Cauchy's integral theorem with the path $\Gamma$ shown in Fig. \ref{Gamma}, the following equation holds:
\begin{figure}
    \centering
    \includegraphics[scale=0.5]{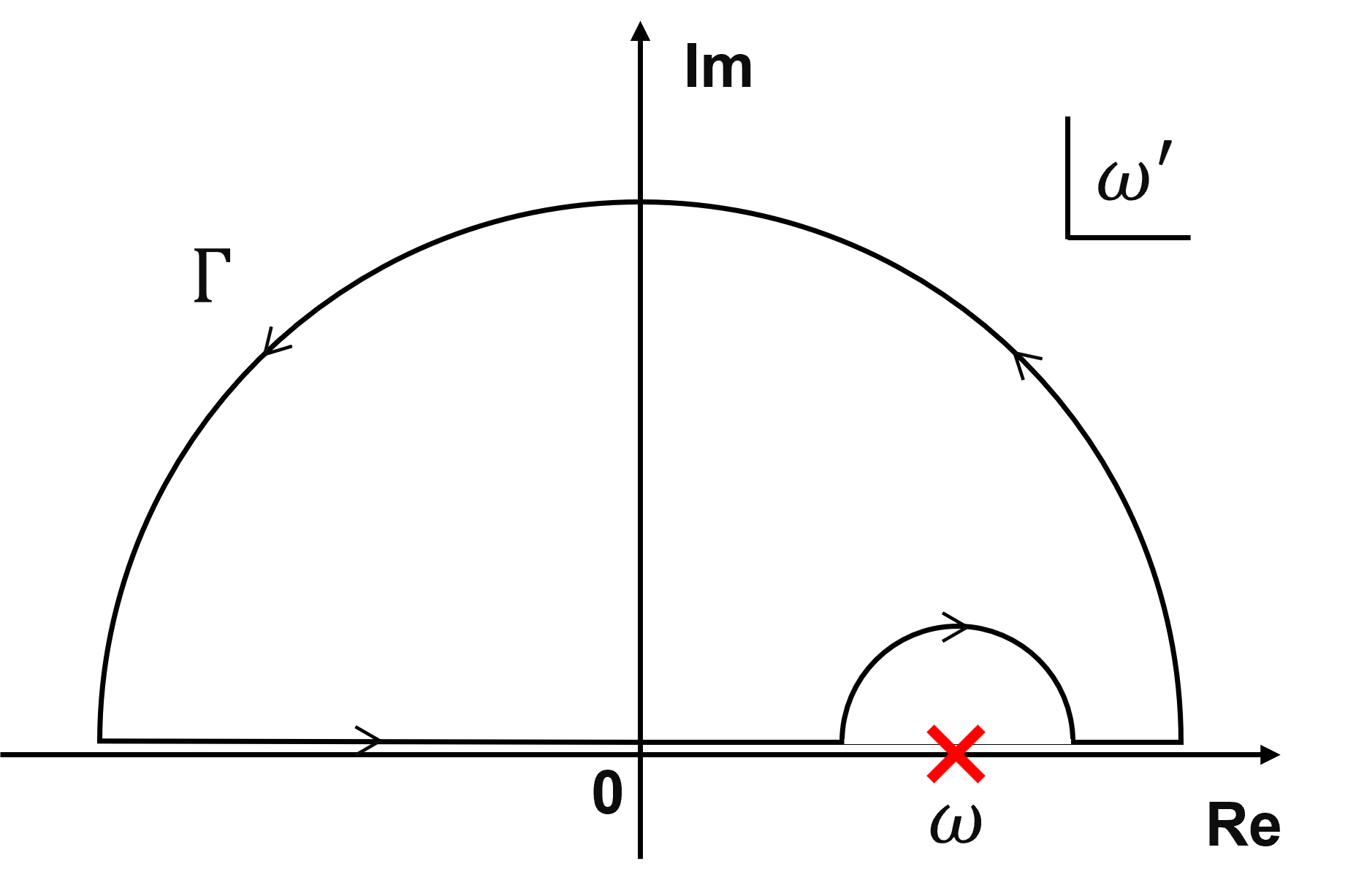}
    \caption{Path of the complex integral. The small semicircle's contribution is the residue at $\omega'=\omega$, and the larger semicircle's contribution is 0.}
    \label{Gamma}
\end{figure}
\be
G(\omega)=\frac{1}{\pi i}\dashint_{-\infty}^{\infty}\frac{d{\omega'}}{\omega'-\omega}G(\omega')
\ee
or
\be
F(\omega)=1+\frac{\omega}{\pi i}\dashint_{-\infty}^{\infty}\frac{d{\omega'}}{\omega'-\omega}\frac{F(\omega')-1}{\omega'},\label{K-K1}
\ee
where 
\be
\dashint_{-\infty}^{\infty}\equiv\lim_{\epsilon\rightarrow0}\qty(\int_{-\infty}^{\omega - \epsilon}+\int_{\omega+\epsilon}^{\infty})
\ee
denotes Cauchy's principal value. Eq. (\ref{K-K1}) is the K-K relation in GL, and this is the relation between the real and imaginary parts of $F(\omega)$:
\ben
\Re F(\omega)&=&1+\frac{\omega}{\pi}\dashint_{-\infty}^{\infty}\frac{d{\omega'}}{\omega'-\omega}\frac{\Im F(\omega')}{\omega'},\label{K-K2}\\
\Im F(\omega)&=&-\frac{\omega}{\pi}\dashint_{-\infty}^{\infty}\frac{d{\omega'}}{\omega'-\omega}\frac{\Re F(\omega')-1}{\omega'}.\label{K-K3}
\een
These two are equivalent, and the K-K relation states that the real and imaginary parts must be related to make the system causal. 

In order to follow the notation used in the literature,
let us define $K(\omega)$ and $S(\omega)$ by\footnote{$K$ and $S$ coincide with the magnification and phase shift when $K,S\ll 1$ \cite{Mizuno:2022xxp}. 
Away from the weak lensing regime, $K$ and $S$ just represent the real and imaginary parts of $F$, respectively.}
\ben
\Re F(\omega)&\equiv&1+K(\omega),\\
\Im F(\omega)&\equiv&S(\omega).
\een
In terms of these quantities, the K-K relation becomes
\ben
&&\frac{K(\omega)}{\omega}=\frac{1}{\pi}\dashint_{-\infty}^{\infty}\frac{d\omega'}{\omega'-\omega}\frac{S(\omega')}{\omega'},\label{K-K4}\\
&&\frac{S(\omega)}{\omega}=-\frac{1}{\pi}\dashint_{-\infty}^{\infty}\frac{d\omega'}{\omega'-\omega}\frac{K(\omega')}{\omega'}.\label{K-K5}
\een

\subsection{Confirmation of the Kramers-Kronig relation for some examples}
In this subsection, we consider two examples to confirm 
that the amplification factor, whose analytic expression is known in the literature, actually obeys the K-K relation derived above.
The formulas for the amplification factor used in this section are all based on the thin-lens approximation. This is the approximation that the lensing effect occurs only in a single plane, and its validity was confirmed in Ref.\cite{Suyama:2005mx}.

\subsubsection{A point-mass lens}
\begin{figure}
    \centering  \includegraphics[scale=0.85]{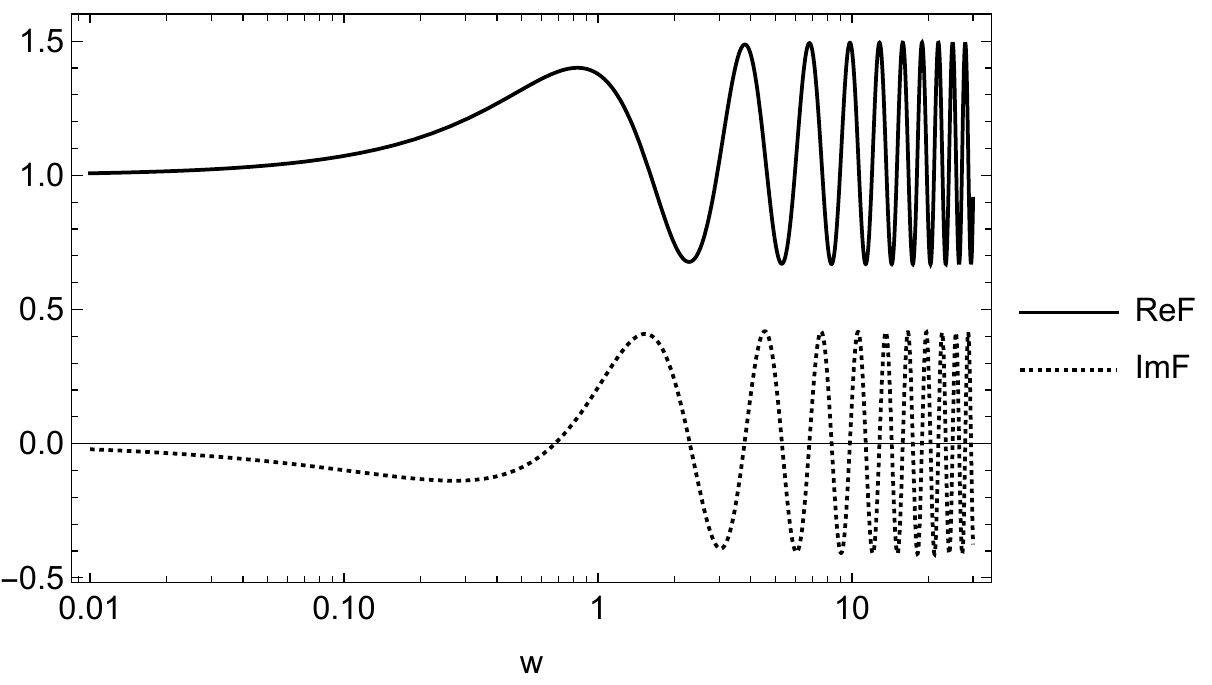}
    \caption{$\Re F$ (solid) and $\Im F$ (dashed) in the case of point mass lens with $y=1$.}
    \label{AbsF}
\end{figure}

The first example is the point-mass lens.
The analytic expression of $F(\omega)$ is given by\cite{schneider1999gravitational}
\be
F(\omega)=\exp\qty[\frac{\pi w}{4}+\frac{iw}{2}\qty(\ln\qty(\frac{w}{2})-2\tau_{min})]\Gamma\qty(1-\frac{iw}{2}){}_1F_{1}\qty(\frac{iw}{2},1;\frac{iwy^2}{2}),\label{ex1}
\ee
where $w\equiv4GM\omega$ with the lens mass $M$ and $y$ is the impact parameter, and $\tau_{min}$ is the dimensionless time delay
\be
\tau_{min} = \frac{2}{(y+\sqrt{y^2+4})^2}-\ln\qty(\frac{y+\sqrt{y^2+4}}{2}).
\ee
It is obvious from Fig.~\ref{AbsF} that $F(0)=1$ and $|F(\infty)|<C$. Besides, $F(u+iv)$ has poles at $w=-2ni$ ($n=1,2,\dots$) that come from $\Gamma\qty(1-\frac{iw}{2})$ but does not have any poles in $I_{+}$. There is also a branch cut that comes from $\ln w$, but this must be placed on the lower half of the complex $\omega$-plane in order to satisfy $F(-\omega)=F^{*}(\omega)$
which comes from the condition that the wave in time domain is not complex but real. 
Therefore, $F$ is regular in $I_{+}$ and satisfies the K-K relation.

\subsubsection{Born approximation}
The second example is the weak lensing in which the amplification factor
is computed to linear order in $\Phi$\cite{Takahashi:2005ug}.
In this approximation, $K$ and $S$ are given by (we also use the thin lens approximation for simplicity):
\ben
K(\omega)&=&\int\frac{d^2k_{\perp}}{(2\pi)^2}\frac{\tilde{\Sigma}(\bm{k_{\perp}})}{\Sigma_0}\frac{\sin\qty(r_F^2k_{\perp}^2/2)}{r_F^2k_{\perp}^2/2},\label{K-K9}\\
S(\omega)&=&\int\frac{d^2k_{\perp}}{(2\pi)^2}\frac{\tilde{\Sigma}(\bm{k_{\perp}})}{\Sigma_0}\frac{\cos\qty(r_F^2k_{\perp}^2/2)-1}{r_F^2k_{\perp}^2/2},\label{K-K10}
\een
where $\tilde{\Sigma}(\bm{k_{\perp}})$ is the Fourier transformed surface mass density and $\Sigma_0$ is a constant that has the dimension of surface mass density, and
\be
r_F(\omega)\equiv\sqrt{\frac{r_{l}(r_0-r_{l})}{\omega r_0}}
\ee
is called the Fresnel scale \cite{Macquart:2004sh} with lens position $r=r_{l}$.
Here, we show that Eq. (\ref{K-K10}) can be obtained from Eq. (\ref{K-K9}) by
applying the K-K relation (\ref{K-K5}). 
In preparation for that, we define $\Omega\equiv r_l(r_0-r_l)k_{\perp}^2/2r_0$.
Then using Eq.~(\ref{K-K9}), the K-K relation (\ref{K-K5}) requires that $S(\omega)$
should be given by
\ben
\label{KK-Born-S}
S(\omega)&=&-\frac{\omega}{\pi}\dashint_{-\infty}^{\infty}
\frac{d\omega'}{\omega'-\omega}\frac{1}{\omega'} 
\int\frac{d^2k_{\perp}}{(2\pi)^2}\frac{\tilde{\Sigma}(\bm{k_{\perp}})}{\Sigma_0}\frac{\sin\qty(r_F^2 (\omega')k_{\perp}^2/2)}{r_F^2(\omega')k_{\perp}^2/2} 
\nonumber \\
&=&-\frac{\omega}{\pi} \int\frac{d^2k_{\perp}}{(2\pi)^2}\frac{\tilde{\Sigma}(\bm{k_{\perp}})}{\Sigma_0}\dashint_{-\infty}^{\infty}
\frac{d\omega'}{\omega'-\omega} \frac{1}{\Omega}
\sin\qty(\frac{\Omega}{\omega'}).
\een
Then, using the formula \cite[Eq. (5.129)]{king_2009}
\be
\frac{1}{\pi}\dashint_{-\infty}^{\infty}\frac{d\omega'}{\omega-\omega'}\sin\qty(\frac{\Omega}{\omega'})\nonumber\\
=\mathcal{H}\qty[\sin\frac{1}{t}](\omega/\Omega)\nonumber\\
=\cos(\frac{\Omega}{\omega})-1, 
\ee
where $\mathcal{H}[~]$ denotes the Hilbert transform,
$S(\omega)$ given by Eq.~(\ref{KK-Born-S}) reproduces Eq.~(\ref{K-K10}).
Thus, the K-K relation holds, or in other words, the causality is satisfied
in the Born approximation.

\subsection{Implications of the Kramers-Kronig relation}
In this subsection, we report some implications that directly
follow from the K-K relation.

\subsubsection{Relation between squares}
First, we show a new relation between the square of the real and 
imaginary parts of $F$. 
Substituting Eq. (\ref{K-K5}) to Eq. (\ref{K-K4}), we get
\ben
\frac{K(\omega)}{\omega}&=&-\frac{1}{\pi^2}\dashint_{-\infty}^{\infty}\frac{d\omega'}{\omega'-\omega}\dashint_{-\infty}^{\infty}\frac{d\omega''}{\omega''-\omega'}\frac{K(\omega'')}{\omega''},
\een
then we have
\ben
\int_{-\infty}^{\infty}d\omega\frac{K^2(\omega)}{\omega^2}
&=&-\frac{1}{\pi^2}\int_{-\infty}^{\infty}d\omega\frac{K(\omega)}{\omega}\dashint_{-\infty}^{\infty}\frac{d\omega'}{\omega'-\omega}\dashint_{-\infty}^{\infty}\frac{d\omega''}{\omega''-\omega'}\frac{K(\omega'')}{\omega''}\nonumber\\
&=&\frac{1}{\pi^2}\int_{-\infty}^{\infty}d\omega'\dashint_{-\infty}^{\infty}\frac{d\omega}{\omega-\omega'}\frac{K(\omega)}{\omega}\dashint_{-\infty}^{\infty}\frac{d\omega''}{\omega''-\omega'}\frac{K(\omega'')}{\omega''}\nonumber\\
&=&\int_{-\infty}^{\infty}d\omega'\frac{S^2(\omega')}{\omega'^2},
\een
where we have exchanged the order of integration with respect to $\omega$ and $\omega'$, and used Eq. (\ref{K-K5}) again. Finally, we get the new relation
\footnote{As it is clear from the derivation, 
the relation (\ref{K-K6}) holds not only for the amplification factor but 
also for any other response functions as long as the K-K 
relation of the type (\ref{K-K4}) and (\ref{K-K5}) holds true.}
\be
\int_{-\infty}^{\infty}d\omega\frac{K^2(\omega)}{\omega^2}=\int_{-\infty}^{\infty}d\omega\frac{S^2(\omega)}{\omega^2}.\label{K-K6}
\ee
This relation may become useful in future observations of GL caused 
by the dark matter inhomogeneities. 
In such observations,
measurements of $\langle K^2(\omega)\rangle$ and $\langle S^2(\omega)\rangle$
enable us to determine the matter powerspectrum on sub-galactic scales
and provide a novel avenue to probe small-scale matter fluctuations \cite{Takahashi:2005ug, Oguri:2020ldf}.
In this respect, Eq.~(\ref{K-K6}) can be used in principle to verify the 
correctness of observed $K(\omega)$ and $S(\omega)$ if the measurements 
cover a wide frequency range to allow estimation of both
left and right hand sides of Eq.~(\ref{K-K6}) to a good approximation.

As a final remark of this subsection, 
it is straightforward to show that the relation (\ref{K-K6}) leads to
\be
\label{KS2K}
\int_{-\infty}^{\infty}d\omega\frac{K^2(\omega)+S^2(\omega)}{\omega^2}
=\int_{-\infty}^{\infty}d\omega\frac{K^2(2\omega)}{\omega^2}.
\ee
On the other hand, it was shown in \cite{Inamori:2021tlx} that 
within the Born approximation for GL caused by dark matter fluctuations
the variances of $K$ and $S$ satisfy the universal relation
\be
\label{unire}
\langle K^2(\omega) \rangle +\langle S^2(\omega) \rangle=
\langle K^2(2\omega) \rangle.
\ee
Because of the simplicity and universality of the relation,
it is expected that there is a simple explanation for the relation (\ref{unire})
based on some fundamental physical principles. 
The coincidence between the relation (\ref{unire}) and the ensemble average of the integrand of Eq.~(\ref{KS2K}) may suggest that the causality is partially responsible for the relation (\ref{unire}) to hold.

\subsubsection{Sum rule}
Second, we show the so-called sum rule. The sum rule is known in the field of optics as the relation between the sum over all frequencies of absorption of a medium and its electric density\cite{lucarini2005kramers}. 
To derive the GL version of this, we deform Eq. (\ref{K-K1}) like
\ben
F(\omega)
&=&1+\frac{1}{\pi i}\dashint_{-\infty}^{\infty}d\omega'\qty(\frac{\omega'}{\omega'-\omega}-1)\frac{F(\omega')-1}{\omega'}\nonumber\\
&=&1+\frac{1}{\pi i}\dashint_{-\infty}^{\infty}du\frac{F(\omega+u)-1}{u}-\frac{2}{\pi}\int_{0}^{\infty}d\omega'\frac{S(\omega')}{\omega'},\label{K-K13}
\een
where we have used $F(-\omega)=F^{*}(\omega)$ in the second line. We now consider complex frequency $\omega=u+iv$ and take $u,v\rightarrow+\infty$ limit, introducing the geometric optics limit:
\be
F(\omega)=\sqrt{\mu_1}+\sum_{j=2}^{n}\sqrt{\mu_j}\exp\qty[i(\omega T_{1j}-\pi n_j)]\label{K-K7},
\ee
where $n$ is the number of images, $\mu_j$ is the magnification factor of the j-th image size, $T_{1j}$ is the arrival time difference between the 1st and j-th image, and $n_j$ is some numbers but not important here. 
In $u,v\rightarrow+\infty$ limit, the oscillating parts of Eq. (\ref{K-K7}) are damped, and then $F(\omega)\rightarrow\sqrt{\mu_1}$. Using this result and taking the real part, Eq. (\ref{K-K13}) becomes
\footnote{
The imaginary part becomes $0=0$.
}
\be
\sqrt{\mu_{1}}=1-\frac{2}{\pi}\int_{0}^{\infty}d\omega\frac{S(\omega)}{\omega}.\label{K-K12}
\ee
This is the GL version of the sum rule, whose meaning is that summing up the imaginary part of $F$ for all frequencies yields the magnification factor of the earliest arriving image.
We expect that this relation could be used as a consistency check in future observations.

\section{Application of the Kramers-Kronig relation}
\label{sec:application}
\subsection{Method for determination of the amplification factor}
In this section, we investigate a potential application of the K-K relation 
to observations of GL of GWs. 
In observations of GWs, what we directly observe is the lensed waveform 
and the separation of the observed waveform into the unlensed waveform
and the amplification factor requires additional procedures.
One approach is to employ the unlensed waveform  
based on templates of some typical sources characterized by the source parameters
and to determine the best fit parameters
\cite{Dai:2018enj}.
In this case, the amplification factor is determined by dividing the measured
waveform with the unlensed template.
However, if the template is determined incorrectly, the obtained amplification factor will also
be different from the true one. 
Such an error will occur, for example, when the source is the precessing binary stars\cite{Apostolatos:1994mx,Chatziioannou:2016ezg} 
since the waveform of the unlensed precessing binary and the microlensed unprecessing binary are very similar \cite{Kim:2023scq}. 
Hence, one may mistake the effect of precession for that of GL, 
resulting in the wrong amplification factor. 
Here we first assume the ideal situation with no measurement error and argue 
that the K-K relation, in principle, can tell us whether the measured amplification factor is truly due to GL or not. Then measurement error is discussed at the end of this subsection.
Again, the analysis assumed in the following discussion does not use a template for the amplification factor, but only for wave sources. Therefore, when we simply refer to "template" below, we are referring to the template for wave sources.

If the true unlensed waveform is $\phi_0$ but we mistakenly employ a 
different template $\hat{\phi}_0$, we obtain the incorrect amplification factor $\hat{F}$ given by
\be
\hat{F}=\frac{\phi_0}{\hat{\phi_0}}F,
\ee
where $F$ is the correct one that satisfies the K-K relation. 
Now, we demonstrate that $\hat{F}$, in general, does not satisfy the K-K relation. 
To this end, we decompose $\hat{F}$ as $\hat{F}=F+\delta F$ and 
assume an extreme case where 
the error $\delta F$ only occurs in $\omega_1\leq \omega\leq\omega_2$\footnote{This assumption is reasonable, at least in the case of the precessing binary. Because precession is the post-Newtonian correction \cite{Apostolatos:1994mx}, it is not negligible only at the end of the inspiral phase, which means that only the high-frequency region is modulated.}:
\be
\delta F(\omega)
\begin{cases}
\ne0&\omega_1\leq\omega\leq\omega_2 \\
=0  &\omega<\omega_1,~\omega_2<\omega
\end{cases}
.
\ee
Substituting $\hat{F}$ for both sides of Eq.~(\ref{K-K1}) and using
that $F$ satisfies the K-K relation, we have
\ben
(LHS)-(RHS)&=&
\hat{F}(\omega)-1-\frac{\omega}{\pi i}\dashint_{-\infty}^{\infty}\frac{d\omega'}{\omega'-\omega}\frac{\hat{F}(\omega')-1}{\omega'}\nonumber \\
&=&\delta F(\omega)-\frac{\omega}{\pi i}\qty(\dashint_{-\omega_2}^{-\omega_1}+\dashint_{\omega_1}^{\omega_2})\frac{d\omega'}{\omega'-\omega}\frac{\delta F(\omega')}{\omega'}.\label{App3}
\een
In general, the second term of the second line is nonzero even when $\omega<\omega_1$ 
or $\omega_2<\omega$, for which case the first term is zero by assumption.
Thus the K-K relation must be broken by the misselection of templates.
This shows that testing whether Eq. (\ref{App3}) vanishes or not 
in principle enables us to conclude whether the claimed lensing signal is correct or not
\footnote{If one uses templates not only for the unlensed waveform
but also for the amplification factor based on some particular lens model,
violation of the K-K relation does not appear.}.

In real observations, there is another cause that leads to the violation of
the K-K relation: truncation of the frequency range due to the limited 
sensitivity of GW detectors. 
When the observable frequency range is restricted to
$[\omega_{min},\omega_{max}]$, computation of the integral in the RHS 
of Eq.~(\ref{K-K1}) by using observational data
is possible only when the range of integration is restricted to this range. 
Thus we must limit the integration as 
\be
\dashint_{-\infty}^{\infty}\rightarrow\dashint_{-\omega_{max}}^{-\omega_{min}}+\dashint_{\omega_{min}}^{\omega_{max}}.
\ee
Then, this causes a further violation of the K-K relation in addition to the misselection of templates. Based on this observation, we introduce a quantity that is a measure of the violation of the K-K relation as
\ben
\Delta(\omega)
&\equiv&\hat{F}(\omega)-1-\frac{\omega}{\pi i}\qty(\dashint_{-\omega_{max}}^{-\omega_{min}}+\dashint_{\omega_{min}}^{\omega_{max}})\frac{d\omega'}{\omega'-\omega}\frac{\hat{F}(\omega')-1}{\omega'}\nonumber\\
&=&\Delta_{tem}(\omega)+\Delta_{tr}(\omega),
\een
where
\be
\Delta_{tem}(\omega)\equiv\delta F(\omega)-\frac{\omega}{\pi i}\qty(\dashint_{-\omega_{max}}^{-\omega_{min}}+\dashint_{\omega_{min}}^{\omega_{max}})\frac{d\omega'}{\omega'-\omega}\frac{\delta F(\omega')}{\omega'},
\ee
and
\ben
\Delta_{tr}(\omega)&\equiv&\frac{\omega}{\pi i}\qty(\int_{-\omega_{min}}^{\omega_{min}}+\int_{-\infty}^{-\omega_{max}}+\int_{\omega_{max}}^{\infty})\frac{d{\omega'}}{\omega'-\omega}\frac{F(\omega')-1}{\omega'}\nonumber\\
&=&\frac{2\omega}{\pi i}\qty(\int_{0}^{\omega_{min}}+\int_{\omega_{max}}^{\infty})\frac{d{\omega'}}{\omega'^2-\omega^2}\qty[K(\omega')+i\frac{\omega}{\omega'}S(\omega')]
\een
is the contribution of the truncation, 
and we have used $F(-\omega)=F^{*}(\omega)$ in the last line.
Notice that once $\hat{F}$ is given by observation, $\Delta$ can be calculated, but $\Delta_{tem}$ and $\Delta_{tr}$ can not be determined respectively.
Considering the region $\omega_{min} \ll \omega \ll \omega_{max}$, we get
\ben
\Re \Delta_{tr}(\omega)&\simeq&C_1 + C_2\qty(\frac{\omega}{\omega_{max}})^2,\label{App1}\\
\Im \Delta_{tr}(\omega)&\simeq&D_1\frac{\omega_{min}}{\omega}+D_2\frac{\omega}{\omega_{max}},
\een
where
\ben
C_1&=&-\frac{2}{\pi}\int_{0}^{\omega_{min}}d\omega\frac{S(\omega)}{\omega},\label{App2}\\
C_2&=&\frac{2\omega_{max}^2}{\pi}\int_{\omega_{max}}^{\infty}d\omega\frac{S(\omega)}{\omega^3},\\
D_1&=&\frac{2}{\pi\omega_{min}}\int_{0}^{\omega_{min}}d\omega~ K(\omega),\\
D_2&=&-\frac{2\omega_{max}}{\pi}\int_{\omega_{max}}^{\infty}d\omega\frac{K(\omega)}{\omega^2},
\een
are dimensionless constants.
\begin{figure}
    \centering
    \includegraphics[scale=0.85]{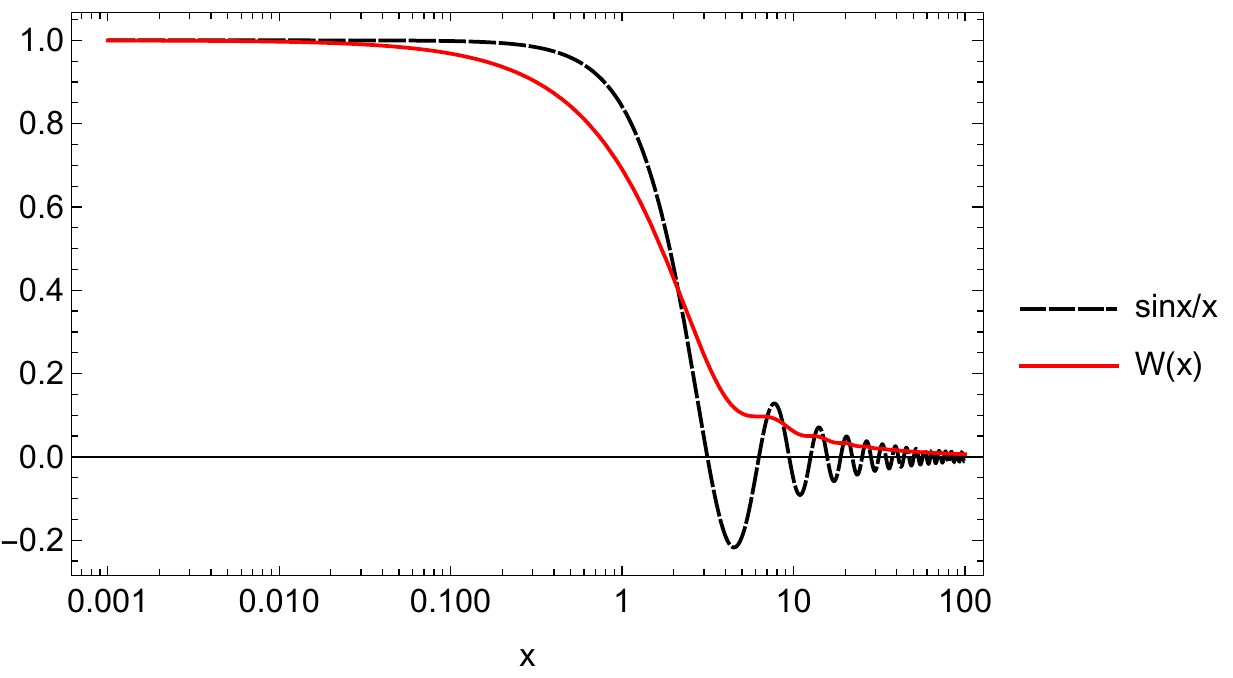}
    \caption{Comparison of $W(x)$ and $\sin x/x$. They have different shapes, but the order coincides in all regions.}
    \label{filter}
\end{figure}
To know the magnitude of $\Delta_{tr}$, we focus on $\Re\Delta_{tr}$
\footnote{If we focus on $\Im\Delta_{tr}$, we have to know the value of $D_1$ and $D_2$. However, we can not use the same method for $D_2$ as for $C_1$, because it is not reasonable to assume that $\omega_{max}$ is in the weak lensing regime and use the Born approximation. This is why we only focus on $\Re\Delta_{tr}$}. 
In the region $\omega_{min}\ll\omega\ll\omega_{max}$, the second term of Eq. (\ref{App1}) is negligible.
Moreover, assuming that $\omega_{min}$ is so small that the weak lensing
is a good description,
we can calculate $C_1$ explicitly by using the Born approximation. 
Substituting Eq.~(\ref{K-K10}) into Eq.~(\ref{App2}) yields
\be
C_1=\int\frac{d^2k_{\perp}}{(2\pi)^2}\frac{\tilde{\Sigma}(\bm{k_{\perp}})}{\Sigma_0}W\qty(r_*^2k_{\perp}^2/2),
\ee
where $r_*\equiv r_F(\omega_{min})$ and
\footnote{$\mathrm{Si}(x)$ is the sine integral defined by $\mathrm{Si}(x)\equiv\int_{0}^{x}\frac{\sin t}{t}dt$.}
\be
W(x)\equiv-\frac{2}{\pi}\qty[\frac{\cos x-1}{x}+\mathrm{Si}(x)-\frac{\pi}{2}].
\ee
Let us compare $W(x)$ with $\sin x/x$, which is the filter function of Eq. (\ref{K-K9}).
As can be seen from Fig. \ref{filter}, 
the orders of magnitude of $W(x)$ and $\sin x/x$ coincides in all $x$.
From above, we can say that
\be
\Re \Delta_{tr}(\omega) \simeq C_1=\mathcal{O}(1)\times K(\omega_{min}), \label{App7}
\ee
which means that from the observed value of $K(\omega_{min})$, 
we can predict how much the K-K relation is violated by the truncation if the claimed amplification factor is correct. 
In this sense, $K(\omega_{min})$ is the threshold for $\Re \Delta(\omega)$. If the observed $\Re \Delta(\omega)$ is sufficiently larger than this value, 
we can assert that the selection of the template is wrong
\footnote{Since $C_1$ has $\mathcal{O}(1)$ uncertainty, if observed $\Re \Delta(\omega)$ is of the same order as $C_1$, it is not possible to make the assertion.}
, by which we may be able to find correct unlensed waveform and the amplification factor. This conclusion remains valid as long as the two assumptions i)$\omega_{min} \ll \omega \ll \omega_{max}$
and ii) weak lensing approximation at the lowest frequency $\omega_{min}$ are simultanesouly satisfied.

What if $\omega_{min}$ is so large and in the geometric optics regime that the Born approximation is no longer valid? In this case, by using Eqs. (\ref{K-K7}) and (\ref{K-K12}), Eq. (\ref{App2}) becomes
\ben
C_1&=&\sqrt{\mu_1}-1+\sum_{j=2}^{n}\frac{2\sqrt{\mu_j}}{\pi}\int_{\omega_{min}}^{\infty}d\omega\frac{\sin(\omega T_{1j}-\pi n_j)}{\omega}\nonumber\\
&=&\sqrt{\mu_1}-1+\sum_{j=2}^{n}\frac{2\sqrt{\mu_j}}{\pi}\frac{\cos (\omega_{min}T_{1j}-\pi n_j)}{\omega_{min} T_{1j}}+\mathcal{O}\qty(\frac{1}{(\omega_{min}T_{1j})^2}),\label{App4}
\een
thus $C_1 = \mathcal{O}(\sqrt{\mu_1}-1)$. 
Therefore, similar to the discussion under Eq. (\ref{App7}), we can warn of the incorrectness of the amplification factor, if the calculated $\Delta$ is sufficiently larger than the observed value of $\sqrt{\mu_1}$. However, we must be careful when $\hat{F}$ is given by the form of Eq. (\ref{K-K7}), but with parameters different from the true ones:
\be
\hat{F}(\omega)=\sqrt{\mu_1'}+\sum_{j=2}^{n'}\sqrt{\mu_j'}\exp\qty[i(\omega T_{1j}'-\pi n_j')].\label{App5}
\ee
In this case, since this $\hat{F}$ itself can be regarded as high-frequency limit of a physically sensible amplification factor that satisfies the K-K relation, the K-K relation is violated only by the truncation, and therefore $\Delta=\mathcal{O}(\sqrt{\mu'_1}-1)$. Thus, we can not report an error in $\hat{F}$ from the violation of the K-K relation
\footnote{If we take the quasi-geometric optics corrections\cite{Takahashi:2004mc} into account, correction terms in Eq. (\ref{App4}) start with $\mathcal{O}(1/\omega)$. However, even in this case, $C_1 = \mathcal{O}(\sqrt{\mu_1}-1)$. It also does not change the fact that $\hat{F}$ of Eq. (\ref{App5}) satisfies the K-K relation. Thus the discussion here remains the same.}. This result can be interpreted as the loss of information obtained from the K-K relation due to the absence of the wave effect.

In addition, it is worth emphasizing that the all above discussion in this subsection is based on the assumption $\omega_{min}\ll\omega_{max}$. If this assumption is not satisfied, the K-K relation would be severely broken by lack of information, as can be seen from the fact that the second term of Eq. (\ref{App1}) is no longer negligible. 
 It should be also noted that $\omega_{max}$ can be in any regime as long as $\omega_{min}\ll\omega_{max}$ is satisfied.

Furthermore, in the realistic case, the amplification factor has a statistical error due to detector noise, which is another factor which may degrade the effectiveness of the K-K relation.
Crudely speaking, if the GW waveform is measured with a statistical significance level given by 
a signal-to-ratio ${\rm SNR}$, the measurement error of the amplification factor will be at the 
level of $\mathcal{O}(1/\rm{SNR})$\cite{Dai:2018enj}.
This uncertainty will propagate into $\Delta$ and give an additional contribution to the violation
of the K-K relation by an amount $\mathcal{O}(1/\rm{SNR})$.
Thus the threshold for $\Delta$ to falsify the claimed source template and the lensing signal
should be set to $\max(C_1, 1/\rm{SNR})$.

Finally, it is important to note that the above discussion does not depend on what the GWs source and lensing object are.

\subsection{Example: A point mass lens}
\begin{figure}
    \centering
    \includegraphics[scale=0.85]{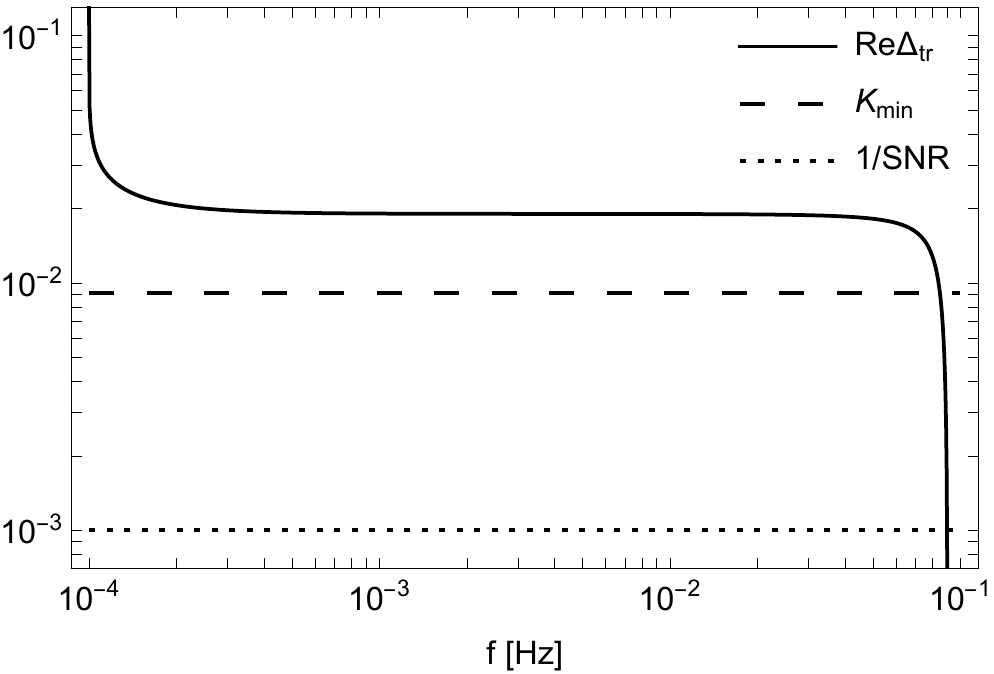}
    \caption{Violation of the K-K relation caused by the truncation $\Re \Delta_{tr}$ (solid). The parameters are $M=10^{6}M_{\odot}$, $y=1$ and the frequency range is decided by $f_{min}=10^{-4}~$Hz and $f_{max}=10^{-1}~$Hz. We also plot $K_{min}$ (dashed) and 1/SNR (dotted).}
    \label{Del_tr}    
\end{figure}
As a demonstration of the methodology described above for general lensing objects,
we show how much the K-K relation is violated by the truncation and statistical error
in the case of the point mass lens for which the amplification factor
is given by Eq.~(\ref{ex1}). 
 As a realistic setting, here we consider LISA. Then we set the observable frequency range as $f_{min}=10^{-4}~\rm{Hz}$, $f_{max}=10^{-1}~\rm{Hz}$. As a lens object, we consider SMBH with $M=10^{6}M_{\odot}$ and $y=1$. In this case, $f_{min}$ is in the week lensing regime because $w_{min}=8\pi GMf_{min}\simeq10^{-2}$. We also assume $\rm{SNR}=10^{3}$ because this is SNR of the BH binary source with $M=10^{6}M_{\odot}$, a typical target for LISA\cite{Babak:2021mhe}.
In Fig. \ref{Del_tr}, we plot $\Re\Delta_{tr}$ with $K_{min}\equiv K(\omega_{min})$ and $1/\rm{SNR}$. From this figure, we can see that Eq. (\ref{App7}) holds except for frequencies close to $f_{min}$
or $f_{max}$. Also, in this typical example, SNR is sufficiently smaller than $K_{min}$, and thus the threshold for $\Re\Delta$ should be set to $K_{min}\simeq0.01$. Therefore, if the template error is greater than a few percent, we can falsify such template.

\section{Conclusion}
It is known that the Kramers-Kronig relation holds true when any system
under consideration respects the causality that output comes only after input. 
We showed that the signal of the gravitational lensing obeys the causality:
waves from a distant source, which
propagate in the gravitational potential created by the lensing objects during
their journey,
never arrive earlier than the null geodesics emitted from the same source simultaneously. 
Inspired by the fact that gravitational lensing has such causality, 
we showed that the Kramers-Kronig relation holds for the amplification factor $F$. 
Since this is a completely new attempt, there are some interesting implications that have not been mentioned in the literature. 
One of them is Eq.~(\ref{K-K6}), which is expected to be used in observations of 
gravitational lensing caused by the dark matter inhomogeneities. 
And the other is the sum rule Eq.~(\ref{K-K12}) which relates the integral of the imaginary part of the amplification factor with the magnification of the first arrival image in geometrical optics.

We also proposed the potential application of the Kramers-Kronig relation 
to observations of gravitational lensing of GWs. 
To determine the amplification factor correctly, 
we need to use the correct template of the GWs source. 
We argued that the false selection of templates can be detected by examining 
the violation of the K-K relation and also calculated the limit of this detection due to the truncation of the frequency range caused by the detector's sensitivity.
Our work suggests that examining the violation of the Kramers-Kronig relation 
may be used for correctly extracting the lensing signal
in the gravitational wave observations.

\section*{Acknowledgements}
We would like to thank Morifumi Mizuno and Ryuichi Takahashi for useful comments and discussions.
This work is supported by the MEXT KAKENHI Grant Number 17H06359~(TS), JP21H05453~(TS),  
and the JSPS KAKENHI Grant Number JP19K03864~(TS).

\bibliography{ref}

\end{document}